\documentclass[11pt,twoside]{article}
\usepackage{amsmath}
\usepackage{amssymb}
\usepackage{cite}
\usepackage{amsmath}
\usepackage{amsfonts}
\usepackage{graphicx}
\usepackage{epsfig}
\usepackage{color}
\usepackage{xspace}

\newcommand{\ket}[1]{\ensuremath{\left| #1 \right\rangle}}

 \newcommand{\EX}[1] {\ensuremath{\left\langle #1 \right\rangle}}
 
\newcommand{\ihbar}{\ensuremath{\frac{i}{\hbar}}}
\newcommand{\half} {\ensuremath{\frac{1}{2}}}

\newcommand{\Fig}[1]{Fig.~\ref{#1}}

\newcommand{\be}{\begin{equation}}
\newcommand{\ee}{\end{equation}}

\newcommand{\bea}{\begin{eqnarray}}
\newcommand{\eea}{\end{eqnarray}}
\newcommand{\ba}{\begin{array}}
\newcommand{\ea}{\end{array}}

\newcommand{\bc}{\begin{center}}
\newcommand{\ec}{\end{center}}

\label{sh}

\begin{document}

\begin{center} {\Large \bf
\begin{tabular}{c}
Persistent entanglement in two coupled SQUID rings\\[-1mm]
in the quantum to classical transition\end{tabular}
 } \end{center}

\bigskip

\bigskip

\begin{center} {\bf
M.J.~Everitt$^{*}$
}\end{center}

\medskip

\begin{center}
{\it
Department of Physics, Loughborough University,
              Loughborough, Leics LE11 3TU, United Kingdom
and
The Centre for Theoretical Physics,
             The British University in Egypt,
             El Sherouk City, Postal No. 11837, P.O. Box 43, Egypt.
}
\smallskip

$^*$Corresponding author e-mail:~~~m.j.everitt@physics.org\\
\end{center}

\begin{abstract}\noindent
We explore the quantum-classical crossover of two coupled, identical, superconducting quantum interference device (SQUID) rings. The motivation for this work is based on a series of recent papers. In ~\cite{1} we showed that the entanglement characteristics of chaotic and periodic (entrained) solutions 
of the Duffing oscillator differed significantly and that in the classical limit entanglement was preserved only in the chaotic-like solutions. However, Duffing oscillators are a highly idealised toy system. Motivated by a wish to explore more experimentally realisable systems we extended our work in ~\cite{2,3} to an analysis of SQUID rings. In ~\cite{3} we showed that the two systems share a common feature. That is, when the SQUID ring's trajectories appear to follow (semi) classical orbits entanglement persists. Our analysis in ~\cite{3} was restricted to the quantum state diffusion unravelling of the master equation - representing unit efficiency heterodyne detection (or ambi-quadrature homodyne detection). Here we show that very similar behaviour occurs using the quantum jumps unravelling of the master 
equation. Quantum jumps represents a discontinuous photon counting measurement process. Hence, the results presented here imply that such persistent entanglement is independent of measurement process and that our results may well be quite general in nature.
\end{abstract}

\section{Introduction}

In  this work  we extend the results of a recent paper~\cite{3}  where we investigated the entanglement   properties  associated   with  the   quantum  classical
crossover of two coupled superconducting quantum interference device (SQUID) rings (comprising of a thick ring enclosing a Josephson junction). 
Here we present a small but significant extension of the series of papers~\cite{1,2,3} which forms a small part of a much larger body of of work - or example see~\cite{bha2,Per98,Hab98,brun97,Bru96,Spi94,Gis93,Gis93b}). In order to avoid too much repetition of text please see~\cite{Per98,1,2,3} and references therein for a more detailed introduction to the subject. Here we present a brief summary of~\cite{3} and our result.

In ~\cite{3} we demonstrated that two coupled SQUID ring's can exhibit entanglement that persists even in the correspondence limit. In order to obtain these trajectories we used the quantum state diffusion unravelling of the master equation and followed a strategy that has seen a lot of success with classically chaotic systems~\cite{Per98}. However - there are an infinite number of ways to unravel the master equation. Hence, a natural concern that arises is that this result might be unravelling dependent. Here we show that very similar behaviour occurs using the quantum jumps unravelling of the master equation. Quantum jumps represents a discontinuous photon counting measurement process. 

Here our interest lay in understanding
how  the  quantum  mechanical phenomena  of entanglement  would
change  as the  coupled  system approached  the  classical limit.   We
showed   \emph{``that   the   entanglement  characteristics   of   two
  `classical'   states  (chaotic   and   periodic  solutions)   differ
  significantly  in the  classical limit.  In particular,  we show[ed]
  that significant  levels of entanglement  are preserved only  in the
  chaotic-like  solutions''}\cite{1}. In~\cite{3}   we extended this investigation to study  the  entanglement
characteristics  in   the  quantum-classical  crossover of two identical coupled SQUID rings.

The correspondence principle in quantum mechanics is usually expressed
in  the  form: \emph{``For  those  quantum  systems  with a  classical
  analogue,  as  Planck's   constant  becomes  vanishingly  small  the
  expectation  values  of  observables  behave  like  their  classical
  counterparts''}\cite{Mer98}.  for SQUID rings such an expression turns out to be problematic and we find that an alternative expression is more appropriate [\emph{sic}]:
\emph{``Consider $\hbar$ fixed (it is) and scale the Hamiltonian so that when
compared with the minimum area $\hbar/2$ in phase space:
(a) the relative motion of the expectation values of the 
	           observable become large and
(b) the state vector is localised.
Then, under these circumstances, expectation values of operators will
behave like their classical counterparts''}\cite{2}.

In order to achieve localisation and model a dissipative chaotic-like system in its correspondence limit we need to introduce decoherence in the right way. Quantum state diffusion has proved particularly successful in many studies of non-linear system. Here we have an It\^{o} increment equation for the state vector of the form~\cite{Gis93,Gis93b}
\begin{eqnarray}\label{eq:qsd}
  \ket{d\psi}  &=& -\ihbar \hat{H}_{sys} \ket{\psi} dt
      +\sum_{j}\left[  \EX{\hat{L}_{j}^{\dagger}} \hat{L}_{j}-\half \hat{L}_{j}^{\dagger}\hat{L}_{j}-\half   
   \EX{\hat{L}_{j}^{\dagger}} \Bigl\langle \hat{L}_{j}\Bigr\rangle \right]  \ket{\psi} dt\nonumber\\
   &&  +\sum_{j}\left[  \hat{L}_{j}-\EX{\hat{L}_{j}} \right]  \ket{\psi} d\xi
\end{eqnarray}
here $\hat{L_j}=\sqrt{2\zeta}\hat{a_j}$,    where   $a_j$    is
the annihilation operator and $dt$ and the $d\xi$ are
complex             Weiner             increments            satisfying
$\overline{d\xi^2}=\overline{d\xi}=0$        and       $\overline{d\xi
  d\xi^{*}}=dt$\cite{Gis93,Gis93b}  where  the  over-bar denotes  the
average over infinitely many stochastic processes.

QSD, however, is not the only unravelling of the master equation and for our results to be general they should be demonstrated to be independent of this choice. This point may be emphasised by observing that previous 
studies have shown that entanglement can be dependent upon the choice
unravelling~\cite{Nha04}. We therefore now choose another unravelling against which we may check our results. We choose an unravelling that is very different from QSD as it is based on a discontinuous 
photon counting measurement process - rather than a continuous interaction - namely quantum jumps~\cite{Car93,Heg93}.
Again this model takes the form of a 
stochastic  It\^{o} increment equation for the state vector but now of the form
  \begin{eqnarray}
    \label{eq:jumps}
    \left| d \psi \right\rangle
    & = & - \frac{i}{\hbar}H   \left|  \psi \right\rangle dt    - \frac{1}{2}\sum_j \left[L_j^\dag L_j - 
            \left\langle L_j^\dag L_j \right\rangle \right] 
            \left|  \psi \right\rangle dt \nonumber\\
    &   & + \sum_j \left[ \frac{L_j}{\sqrt{\left\langle L_j^\dag L_j \right\rangle}} 
          - 1 \right] \left|  \psi \right\rangle dN_j
  \end{eqnarray}
where  $dN_j$   is  a  Poissonian   noise  process  such   that  $dN_j
dN_k=\delta_{jk}dN_j$,  $dN_j  dt=0$ and  $\overline{dN_j}=\left\langle
  L_j^\dag L_j \right\rangle  dt$, i.e. jumps occur  randomly at a
rate that is determined by $\left\langle L_j^\dag L_j \right\rangle$.

In~\cite{1} we studied the entanglement dynamics (characterised via the entropy  of entanglement $S\left(\rho_{i}\right)
=-\mathrm{Tr}[\rho_{i}\ln\rho_{i}]$) in two
coupled Duffing oscillators\cite{1} (extending one dimensional analysis in, for example,~\cite{Bru96,Per98}). The
Hamiltonian for each oscillator was given by
\begin{equation}
H_{i}=\frac{1}{2}p_{i}^{2}+\frac{\beta^{2}}{4}q_{i}^{4}-\frac{1}{2}q_{i}
^{2}+\frac{g_{i}}{\beta}\cos\left(  t \right)  q_{i}+\frac
{\Gamma_{i}}{2}(q_{i}p_{i}+p_{i}q_{i})
\end{equation}
where $q_{i}$ and $p_{i}$,  $L_{i}=\sqrt{2\Gamma_{i}}  a_{i}$ (for  $i=1,2$),  where
$a_{i}$ is  the annihilation operator. Here $g_{i}=0.3$ and $\Gamma_{i}=0.125$, \cite{Bru96,Per98,1}. In this  work  the
parameter $\beta$  is a scaling  parameter used to generate the correspondence limit.
The Hamiltonian for
the coupled system is:
\begin{equation}\label{eq:hamSys}
H=H_{1}+H_{2}+\mu q_{1}q_{2}
\end{equation}
with  $\mu=0.2$.  

\begin{figure}[!t]
\begin{center}
\resizebox*{0.4\textwidth}{!}{\includegraphics{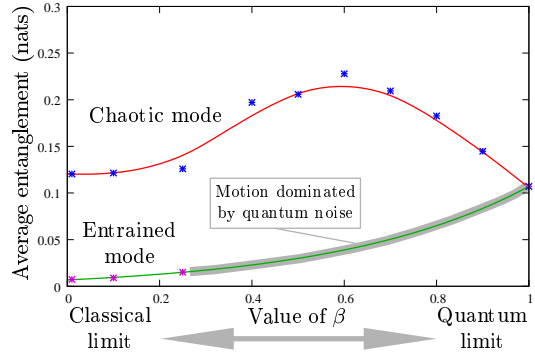}}
\end{center}
\caption{Mean entropy of entanglement as a function of $\beta$ for the chaotic-like
  and periodic  (entrained) states. Here we see  that the entropy
  of entanglement for the system  in the chaotic state does  not vanish as $\beta$
  approaches the classical regime.  (Note: Figure and caption reproduced from ~\cite{1})\label{fig:average.qsd}}
\end{figure}
\begin{figure}[t]
\begin{center}
\resizebox*{0.4\textwidth}{!}{\includegraphics{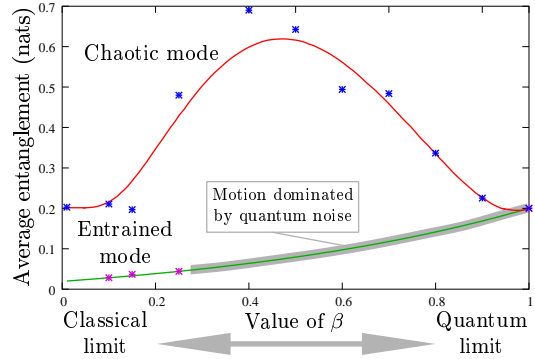}}
\end{center}
\caption{The calculation  of figure~\ref{fig:average.qsd} using quantum 
  jumps instead  of quantum  state diffusion. Again  we show  the mean
  entropy of entanglement as a function of $\beta$  for the chaotic-like
  and periodic (entrained) states.  As with quantum state diffusion we
  see  that when  using  quantum jumps  the  entropy of entanglement  for the
  system in  the chaotic state  does not vanish as  $\beta$ approaches
  the classical regime. (Note: Figure and caption reproduced from ~\cite{1})\label{fig:jumps}}
\end{figure}

The dynamics of the oscillators have two distinct modes of operation; entrained \& periodic and un-entrained \& chaotic.
When the oscillators are  entrained  we found that, as one would expect, the entanglement  falls as the system  approaches the classical regime.  In the un-entrained \& chaotic  mode of operation we found that significant
average entangled was manifest both in the quantum and classical limit. These results are shown in ~\Fig{fig:average.qsd} using quantum state diffusion and~\Fig{fig:jumps} for quantum jumps unravellings of the master equation.

In~\cite{2,3} we extended this investigation to SQUID's.
Here the \emph{``classical''} dynamics are described by the resistively shunted junction (RSJ) model:
\begin{equation}\label{eq:rsj}
C \frac{d^2\Phi}{dt^2}+\frac{1}{R}\frac{d\Phi}{dt}+\frac{\Phi-\Phi_x}{L}+I_c \sin\left(\frac{2\pi \Phi}{\Phi_0}\right)
=I_d \sin \left(\omega_d t \right)
\end{equation}
where $\Phi$ is the magnetic flux contained within the ring $\Phi_x$,  $C$,  $I_c$,  $L$, $R$, $I_d$,  $\omega_d$ and $\Phi_0=h/2e$ are  the
external flux bias, capacitance and critical current of the weak link,
ring inductance, resistance, drive amplitude,
drive frequency and flux quantum, respectively. Here, $C=1\times10^{-13}$F,  $L=3\times10^{-10}$H, $R=100\Omega$, $\beta=2$,
$\omega_d=\omega_0$, $\Phi_x=0.5\Phi_0$ and $I_d=0.9\,\mu\mathrm{A}$.

[\emph{sic}~\cite{3}] ``We  can  then  rewrite   (\ref{eq:rsj})  in  the  standard,  universal
oscillator   like,   form  by   making   the  following   definitions:
$\omega_0=1/\sqrt{LC}$,               $\tau=\omega_0               t$,
$\varphi=(\Phi-\Phi_x)/\Phi_0$, $\varphi_x=\Phi_x/\Phi_0$, $\beta=2\pi
L  I_c/\Phi_0$,  $\omega=\omega_d/\omega_0$, $\varphi_d=I_d  L/\Phi_0$
and   $\zeta=1/2\omega_0    RC$. This yields the following equation of motion:
\begin{equation}\label{eq:rsjNorm}
\frac{d^2\varphi}{d\tau^2}+2\zeta\frac{d\varphi}{d\tau}+\varphi+\frac{\beta}{2\pi} \sin\left[2\pi\left( \varphi+\varphi_x\right)\right]
=\varphi_d \sin \left(\omega \tau \right)
\end{equation}
In  this system of  units we  then see  that we  can scale  the system
Hamiltonian  through   changing  either  $C  \rightarrow   aC$  or  $L
\rightarrow  bL$  so long  as  we  also  make the  following  changes:
$R\rightarrow\sqrt{{b}/{a}}R$,  $I_d\rightarrow  {I_d}/{\sqrt{b}}$ and
$\omega_d\rightarrow {\omega_d}/{\sqrt{ab}}$. \ldots  We change $a$  so that $C$
varies  between  $1\times 10^{-16}$\,F  (quantum  limit) and  $1\times
10^{-9}$\,F  (classical  limit),  changing  other  circuit  parameters
in line with the above methodology.''

The Hamiltonian is:
\begin{equation}\label{eq:qmBase}
\hat{H_i}=\frac{\hat{Q}^2_i}{2C}+\frac{\left(\hat{\Phi}_i-\Phi_{x_i}(t)\right)^2}{2L}-
\frac{\hbar I_c}{2e} \cos\left(\frac{2\pi\hat{\Phi}_i}{\Phi_0} \right)
\end{equation}
with $\left[\hat{\Phi}_i,\hat{Q}_i\right]=i\hbar$.

As usual we define:
$\hat{x}_i=\sqrt{{C\omega_0}/{\hbar}}\hat{\Phi}_i$                  and
$\hat{p}_i=\sqrt{{1}/{\hbar C \omega_0}}\hat{Q}_i$. and $\hat{H}_i'=\hat{H}_i/\hbar \omega_0$ so that
\begin{equation}\label{eq:qmNorm}
  \hat{H}_i'=\frac{\hat{p}^2_i}{2}+\frac{[\hat{x}_i-x_i(t)]^2}{2}-\frac{I_c}{2e \omega_0}\cos \left( \Omega \hat{x}_i\right)
\end{equation}
where $\Omega=\left[(4e^2/\hbar)\sqrt{(L/C)} \right]^{1/2}$.

One further correction to the Hamiltonian is needed to correctly introduce damping~\cite{Per98} which now becomes:
\begin{equation}\label{eq:qmNorm2}
  \hat{H}'_i=\frac{\hat{p}^2_i}{2}+\frac{[\hat{x}_i-x_i(t)]^2}{2}-\frac{I_c}{2 e \omega_0}\cos \left( \Omega   
           \hat{x}_i\right)+\frac{\zeta}{2}\left(\hat{p}_i\hat{x}_i+\hat{x}_i\hat{p}_i\right)
\end{equation}
So, for two coupled SQUID's we have
\begin{eqnarray}\label{eq:HamCoupled}
  \hat{H}_{total}&=&\sum_{i\in\{1,2\}}\left\{\frac{\hat{p}^2_i}{2}+\frac{[\hat{x}_i-x_i(t)]^2}{2}-\frac{I_c}{2 e \omega_0}\cos \left( \Omega   
           \hat{x}_i\right)+\right. \nonumber \\ 
           &&\left.\frac{\zeta}{2}\left(\hat{p}_i\hat{x}_i+\hat{x}_i\hat{p}_i\right)\right\}+\mu \hat{x}_1 \hat{x}_2 \nonumber
\end{eqnarray}
where we have chosen $\mu=0.2$ (as this is the value that we used in~\cite{1}).

\begin{figure}[!tb]
\centerline{\resizebox*{0.8\textwidth}{!}{\includegraphics{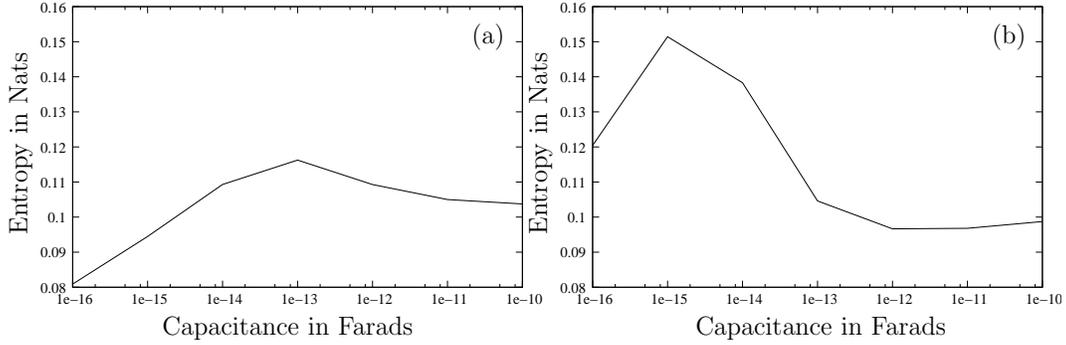}}}
\vspace*{8pt}
\caption{Mean entanglement  entropy as  a function of  Capacitance two
  coupled SQUID rings using  (a) quantum state diffusion and (b) quantum jumps unravellings of the master equation. In both figures we  see that the entanglement entropy for
  system  does  not  vanish   even  as  it  approaches  its  classical
  limit.
  Note: that unlike in~\Fig{fig:average.qsd} and~\Fig{fig:jumps} in this figure the quantum limit is on the left hand side and the classical limit on the right.
  \label{fig:s1}}
\end{figure}
In~\Fig{fig:s1}(a) we show the mean entanglement of the two SQUID rings found by using the Quantum state diffusion unravelling of the master equation (these results were also presented in~\cite{3}).
Here small capacitance is the quantum limit and large capacitance is the correspondence limit. The capacitance was changed via use of the  scaling  parameters   $a$ of the discussion above.  [\emph{sic}~\cite{3}] \emph{``However  we  note  that  the
entanglement entropies  presented here  are is the  average entanglement
over either a long time period or many similar trajectories. It is not
the entanglement associated with  the average density operator taken of
many  experiments.   This  average  entanglement   cannot  therefore  be
considered   usable    in   a   quantum    information   sense.    In
figure~\ref{fig:s1} we show this average entanglement entropy. Here the
averaging  of each  trajectory  was  determined on  a  point by  point
basis.  A sufficient  averaging  was used  so  as to  ensure that  the
results presented here had settled to within a percent or so ... As for  the Duffing oscillators,  here the mean entanglement  does not
appear to  vanish in the classical limit  (large capacitance). Another
surprising  feature in common  with the  Duffing oscillator  results is
that  the average entropy  is not  maximum at  the most  quantum limit
(smallest capacitance).''}

In~\Fig{fig:s1}(b) we present the result of this paper - here we have simply reproduced the calculations of ~\Fig{fig:s1}(a) using the quantum jumps unravelling of the master equation. We note that for the quantum jumps model that - especially in the quantum limit - it takes much longer for the averages to settle to their final values and there is some small error attached to each of the data points. However there is a good qualitative agreement between these results and those obtained for the Duffing oscillator. Is seems then that such persistent entanglement is independent of measurement process and that our results may well be quite general in nature.

\section*{Acknowledgments}
The author  would like  to thank The  Physics Grid  (Loughborough) and
Loughborough University HPC service for use of their facilities.
\bibliographystyle{unsrt}
\bibliography{references}

\end{document}